\documentclass[twocolumn]{emulateapj}
\usepackage{amsmath, amsfonts, amssymb,natbib}
\begin{document}

\title{Polar disk galaxy found in wall between voids}
\author{K. Stanonik\altaffilmark{1}, E. Platen\altaffilmark{2}, M. A. Arag\'on-Calvo\altaffilmark{3}, J. H. van Gorkom\altaffilmark{1}, R. van de Weygaert\altaffilmark{2}, J. M. van der Hulst\altaffilmark{2}, P. J. E. Peebles\altaffilmark{4}}

\altaffiltext{1}{Department of Astronomy, Columbia University, Mail Code 5246, 550 West 120th Street, New York, NY 10027, USA; email: kstanonik@astro.columbia.edu}
\altaffiltext{2}{Kapteyn Astronomical Institute, University of Groningen, PO Box 800, 9700 AV Groningen, the Netherlands}
\altaffiltext{3}{The Johns Hopkins University, 3701 San Martin Drive, Baltimore, MD 21218, USA}
\altaffiltext{4}{Joseph Henry Laboratories, Princeton University, Princeton, NJ 08544, USA}

\begin{abstract}
We have found an isolated polar disk galaxy in what appears to be a cosmological wall situated between two voids.  This void galaxy is unique as its polar disk was discovered serendipitously in an H {\sc i} survey of SDSS void galaxies, with no optical counterpart to the H {\sc i} polar disk. Yet the H {\sc i} mass in the disk is comparable to the stellar mass in the galaxy.  This suggests slow accretion of the H {\sc i} material at a relatively recent time.  There is also a hint of a warp in the outer parts of the H {\sc i} disk.  The central, stellar disk appears relatively blue, with faint near UV emission, and is oriented (roughly) parallel to the surrounding wall, implying gas accretion from out of the voids.  The considerable gas mass and apparent lack of stars in the polar disk, coupled with the general underdensity of the environment, supports recent theories of cold flow accretion as an alternate formation mechanism for polar disk galaxies. 
\end{abstract}


\section{Introduction} \label{intro}

Polar ring galaxies are peculiar galaxies encircled by a ring of stars, gas and dust with perpendicular spin \citep{1994AJ....108..456R}.  These distinct kinematic components are confirmed by observations of their H {\sc i} distribution \citep{1984MNRAS.208..111S, 1987ApJ...314..457V,1997AJ....113..585A}, which generally exhibit extended emission throughout the polar ring with a central gap.  Kinematically, we see that some of these structures are more disk-like than ring-like \citep{2006ApJ...643..200I}, are massive enough to be self gravitating \citep{1994AJ....107..958A}, and have the suggestion of spiral arms \citep{1997AJ....113..585A}.  The central galaxy is generally a rotationally supported S0 or elliptical galaxy and shares a common geometric center with the polar disk, however is relatively gas poor.  Polar disk galaxies are relatively rare, with only a few percent of S0s that have now or have had a polar ring \citep{1990AJ....100.1489W}.  They are particularly useful and interesting as their perpendicular dynamics allow a three dimensional probe of the dark matter potential \citep{1983AJ.....88..909S,1994ApJ...436..629S,1996A&A...305..763C}.

The formation mechanism for polar disks is not entirely clear, but presumably requires a second formation event to form the perpendicular spin components.  Originally, a major merger or close encounter was assumed necessary \citep{1983AJ.....88..909S}, however
neither theory resolves the problem of the central disk retaining its rotational support in systems with significantly massive polar disks, as with NGC 4650A \citep{2003A&A...401..817B,2005A&A...437...69B,2006ApJ...643..200I}.  The issue is also confused by indeterminate ages of observed polar disks, which are partly dependent on the stability of polar oriented structures \citep{2004ASSL..312..273S}.  More recent theories and simulations suggest cold accretion along cosmological filaments may form and continue to feed these polar disks \citep{2006ApJ...636L..25M}.  

We have found a polar disk galaxy in relatively isolated, void-like conditions, situated in a tenuous sheet-like configuration between two voids.  It was found as part of an H {\sc i} survey of 15 void galaxies selected using an early version of the Watershed Void Finder (\citealt{platen2007}) and the Sloan Digital Sky Survey (SDSS, \citealt{Adelmanmccarthy08}), with each galaxy in a different void.  It has no optical polar ring visible in the SDSS imaging but a massive, extended H {\sc i} polar disk.  This suggests cold flow accretion as the most likely formation mechanism, and sheds a new and surprising light on the nature of galaxies populating the most desolate areas in the Universe (cf e.g. \citealt{2001ApJ...557..495P}).  Observations are described in \S \ref{obs}, shown in \S \ref{results}, and comparisons with theory made in \S \ref{dis}.

\begin{figure*}[ht!]
\centering
\includegraphics[height=3.3in, angle=0]{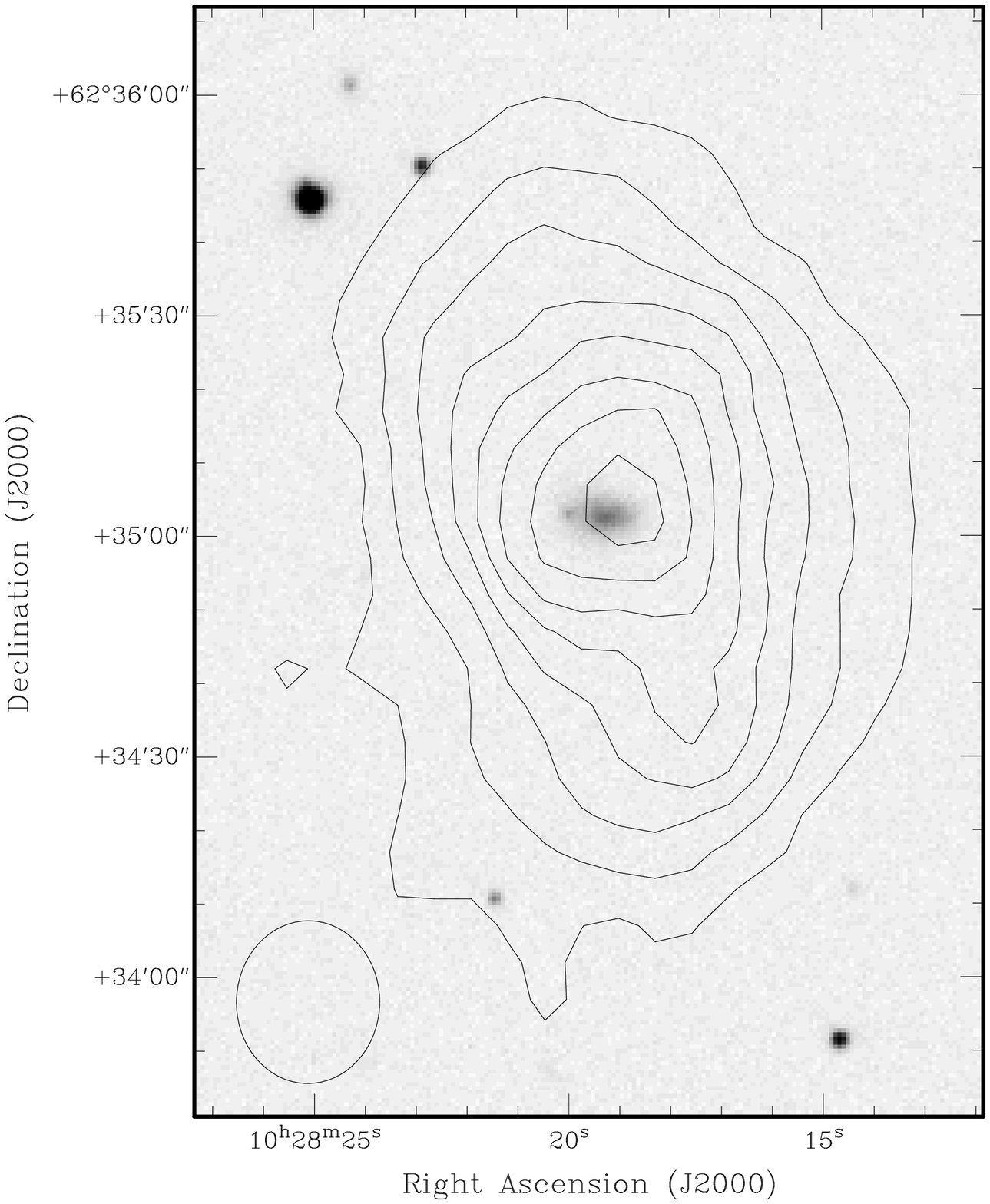}
\includegraphics[height=3.3in, angle=0]{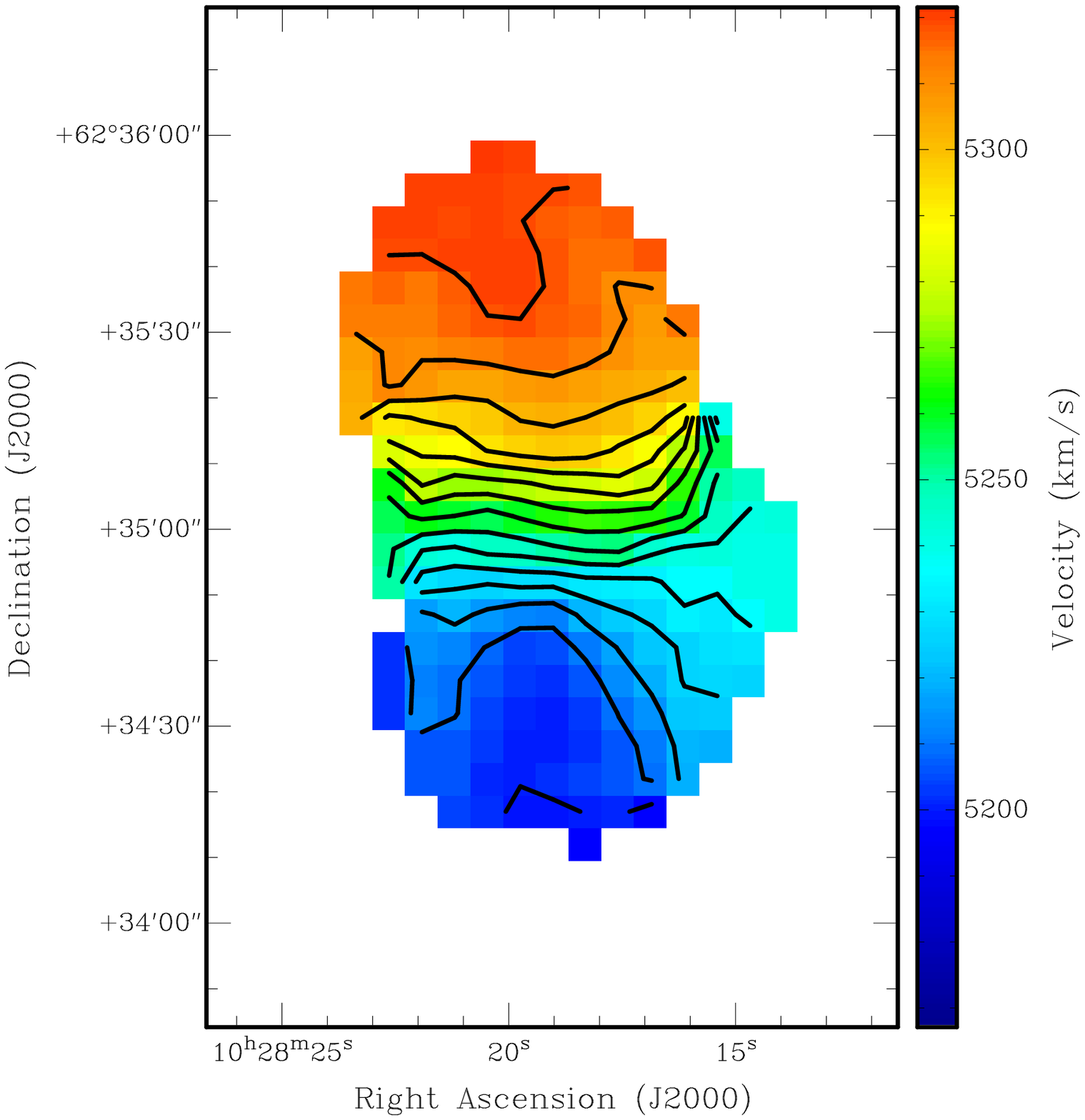}
\caption{SDSS J102819.24+623502.6.  On the left, the g-band image overlaid with H {\sc i} contours.  The column density contours are 0.76 (1.9$\sigma$),   2.0,   3.3,   4.5,   5.9,   7.0,   8.4,   9.6 x 10$^{20}$ cm$^{-2}$.  On the right, the intensity weighted velocity field overlaid with 8.5 km s$^{-1}$ contours.\label{contvels}}
\end{figure*}

\section{Observations} \label{obs}

SDSS J102819.24+623502.6 is part of a pilot for a larger study of galaxy formation in underdense regions of the SDSS redshift survey (Stanonik et al. 2009, in preparation).  Our lonely galaxy is situated at a RA \& Dec (J2000) of 10:28:19.02 +62:35:02.0 and a redshift of z= 0.0178446.  It is fairly blue, with a g-band magnitude of 17.67 and r-band magnitude of 17.41 as reported in the sixth data release from the SDSS.  Note that these images reach to magnitudes of g $\sim$ 23, as the drift-scan mode employed by the SDSS results in an effective exposure time in each band of 54.1 s \citep{2000AJ....120.1579Y}.  A distance of 76 Mpc is calculated assuming H$_o$ = 70 km s$^{-1}$ Mpc$^{-1}$, which gives a scale of 1$^{\prime\prime}$=370 pc.  

The 21 cm line of our target was observed for 12 hours at Westerbork Synthesis Radio Telescope(WSRT) in Maxi-short configuration, which optimizes imaging of very extended sources in a single track observation with shortest baselines of 36, 54, 72 and 90 meters.  We observed in 4 IFs each with 512 channels, a 10 MHz bandwidth, and 2 polarizations.  The target redshift corresponds to an optical heliocentric velocity of 5302 km s$^{-1}$, so channel increments of 19.5 kHz give a velocity resolution of 4.27 km s$^{-1}$.

All calibration was done using standard AIPS procedures. Continuum emission was subtracted from the UV data by linear interpolation from the line free channels.  Image cubes were created with a CLEAN box around the H {\sc i} emission that cleaned down to 2 mJy beam$^{-1}$ (4 $\sigma$).  Images were created with uniform weighting and robust 1 to give a balance of high resolution, a beam size of 22\arcsec.1 x 19\arcsec.5, with low noise, a rms of 0.5 mJy beam$^{-1}$.  This corresponds to a 3$\sigma$ detection limit in column density of $6.2 \times 10^{18}$ cm$^{-2}$ and in mass of $2.0 \times 10^8 M_\sun$ for a galaxy with velocity width 100 km s$^{-1}$.  $0^{th}$ and $1^{st}$ moment maps were made, with 1.5$\sigma$ clipping applied to a mask that is Hanning and Gaussian smoothed over 3 cells, both in velocity and spatially, respectively.

\section{Results} \label{results}

\subsection{Galaxy Parameters}
The total H {\sc i} image and intensity weighted velocity field are shown in Fig. \ref{contvels}.  We calculate a total H {\sc i} mass of M$_{\textrm{H \textsc{I}}}=3.0 \pm 0.5 \times 10^9$ M$_{\sun}$.  There is an angle of 90$^{\circ}$ between the central stellar disk and the polar H {\sc i} disk, which warps at a distance of 20 kpc to 77$^{\circ}$.  

\begin{figure}[b!]
\begin{center}
\includegraphics[height=1.9in, angle=0]{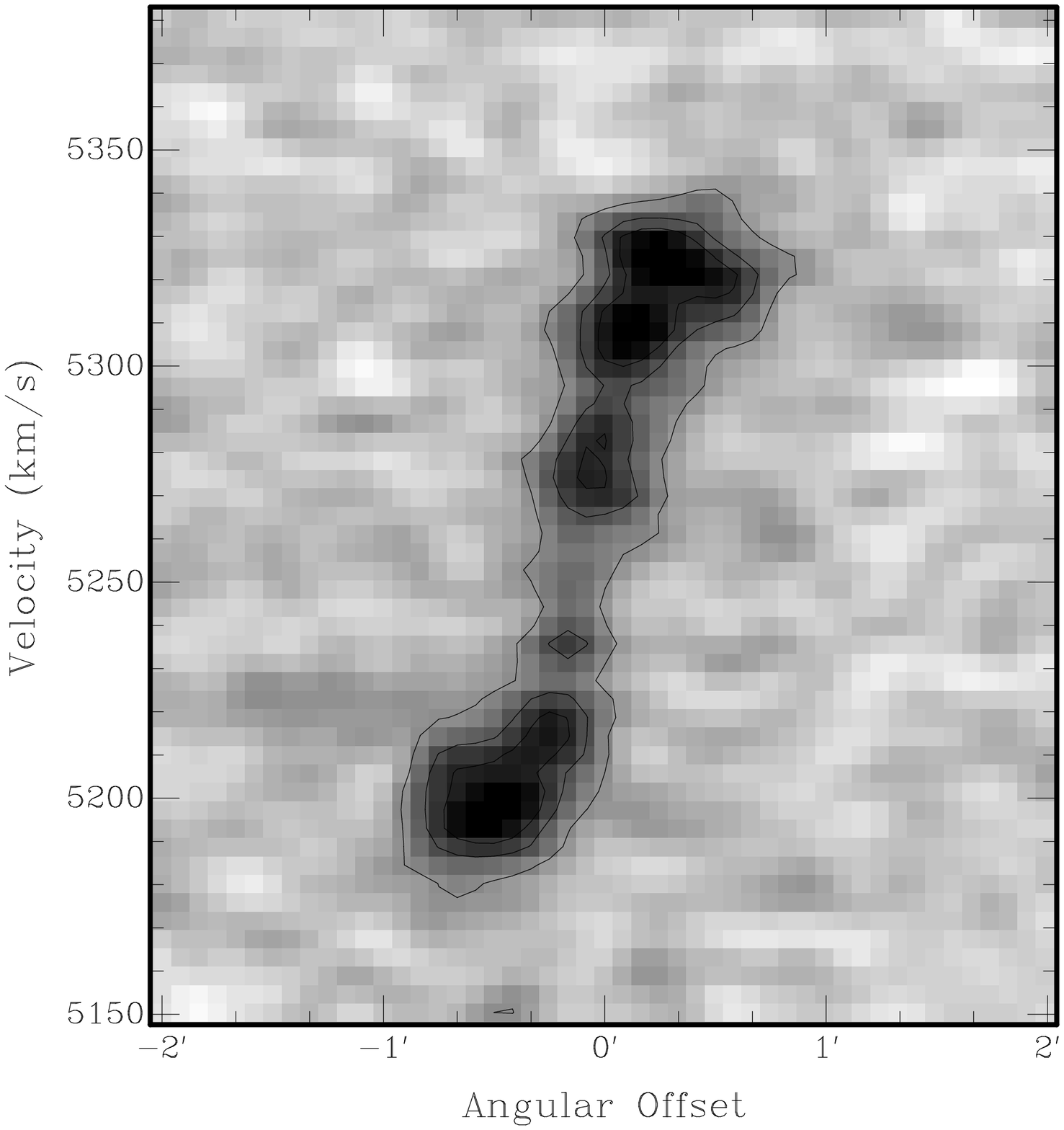}
\includegraphics[height=1.9in, angle=0]{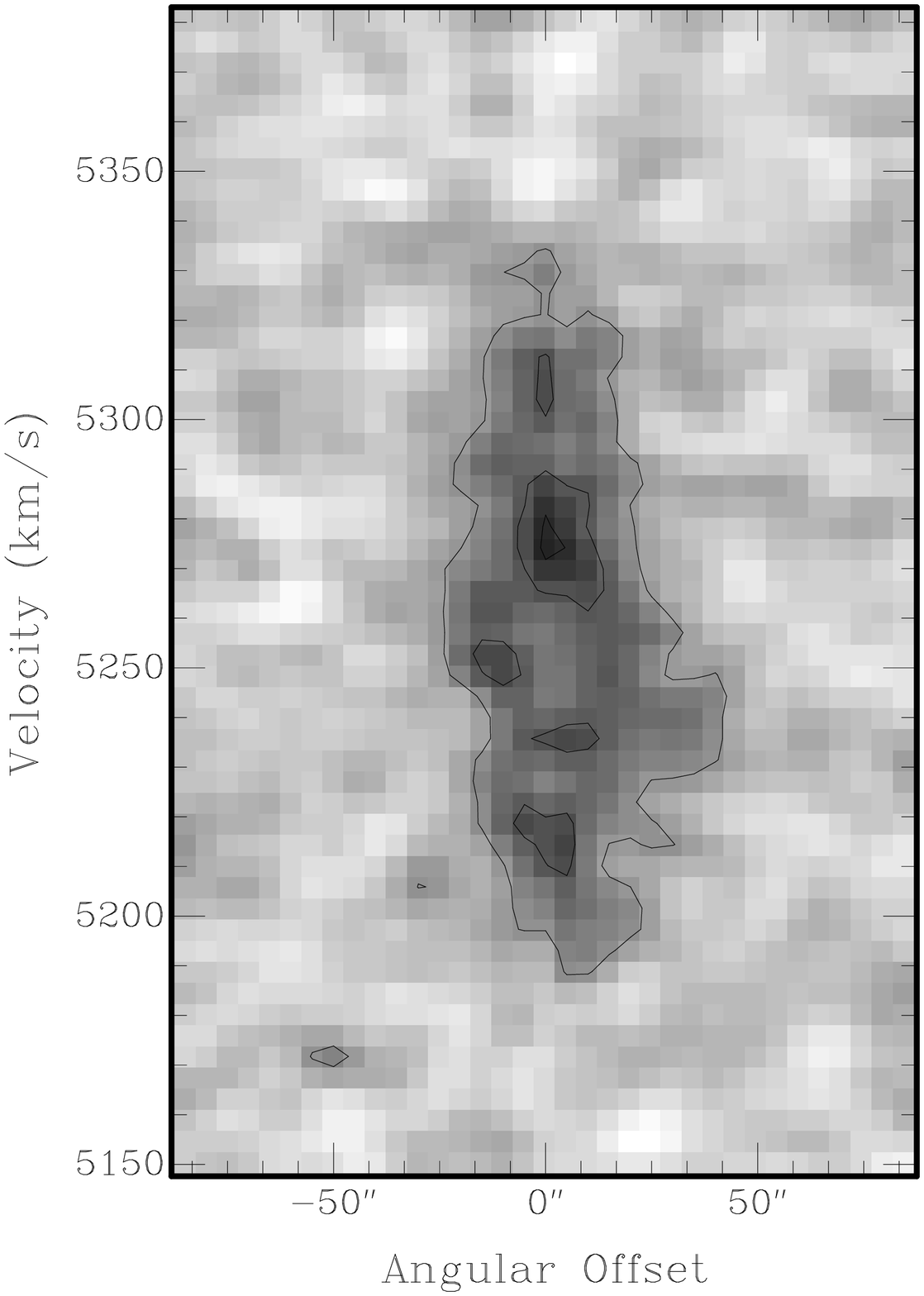}
\caption{PV diagrams aligned with the polar disk (left) and central stellar disk (right).  Note the central hole in the gas distribution in both slices. Contours in both images are at increments of 1.5 mJy beam$^{-1}$ (3$\sigma$).\label{posvel}}
\end{center}
\end{figure}

A comparison of slices through the data cube in position-velocity along the major axis of the H {\sc i} disk compared to the major axis of the central stellar disk (Fig. \ref{posvel}) clearly demonstrate that the regular rotation within this misaligned gas disk is completely decoupled from the stellar component.  The major axis slice (left panel, Fig. \ref{posvel}) confirms the notion from the velocity field that the H {\sc i} is a disk and not a ring. The rotation curve increases non-linearly with increasing radius and appears to flatten out at a radius of about 30\arcsec, or 11 kpc. The rotation at the last measured point (about 22 kpc) is 100 km s$^{-1}$ (assuming an inclination of 40$^\circ$ derived from a best fit of the velocity field). This implies a dynamical mass of 5 x 10$^{10}$ M$_{\sun}$ within 22 kpc. The SDSS $r_{90}$ value of 6.6$^{\prime\prime}$  gives a radius of 2.4 kpc, so the extent of the H {\sc i} gas is about 9 times the extent of the perpendicular stellar disk.  The inverse concentration index for SDSS galaxies, taken from the ratio of $r_{50}$ to $r_{90}$, correlates tightly with morphological type \citep{2001AJ....122.1238S}, and in our central, stellar galaxy is consistent with an exponential, rotationally supported disk.  

Analysis of the optical spectra available from SDSS allows an approximate measure of the stellar mass in the central galaxy. The parameters are calculated by Yip et al. (2009, in preparation) who utilizes the spectrum services provided by \cite{2006IAUSS...3E..76D}, and the stellar population model by \cite{2003MNRAS.344.1000B} to find a mass of $1.6 \times 10^9 M_\sun$ for the part of our galaxy falling within the 3\arcsec ~fiber.  This provides a lower limit for the stellar mass of the entire galaxy.  Since this area is roughly half the optical extent of our tiny galaxy, we expect the total stellar mass to be twice that, roughly equal to the H {\sc i} mass in the polar disk. This is similar to the polar disk simulated by \cite{2006ApJ...636L..25M} which had a baryonic mass in the polar ring of only a third the mass of the central S0 galaxy, and NGC 4650A which has twice the mass in gas and stars in its polar disk as in the central host galaxy \citep{2002AJ....123..195I}.

\begin{figure}[t]
\begin{center}
\includegraphics[width=2.5in, angle=0]{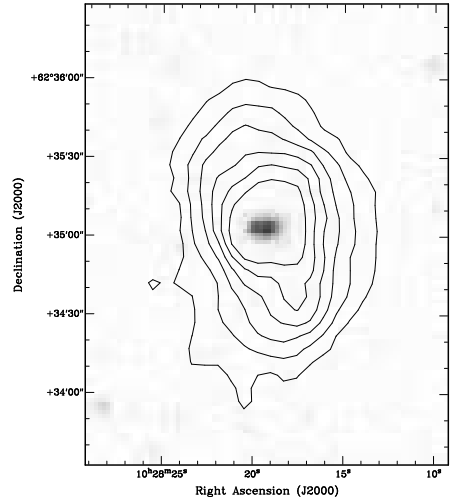}
\caption{SDSS J102819.24+623502.6, galex NUV image overlaid with H {\sc i} contours. H {\sc i} contours are incremented as in Fig. \ref{contvels}, NUV detection is 19.0 magnitude after 1619 second exposure.\label{galex}}
\end{center}
\end{figure}

Our target was also observed by GALEX as part of its All Sky Survey (AIS), and serendipitously in the field of a guest observer 1619 s NUV exposure. Fig. \ref{galex} overplots H {\sc i} contours over the deep NUV image.  The 19th magnitude UV emission follows the disk of the central optical galaxy, with no detection along the polar disk.

\subsection{Large Scale Environment}
\begin{figure}[b!]
\centering
\includegraphics[height=3in]{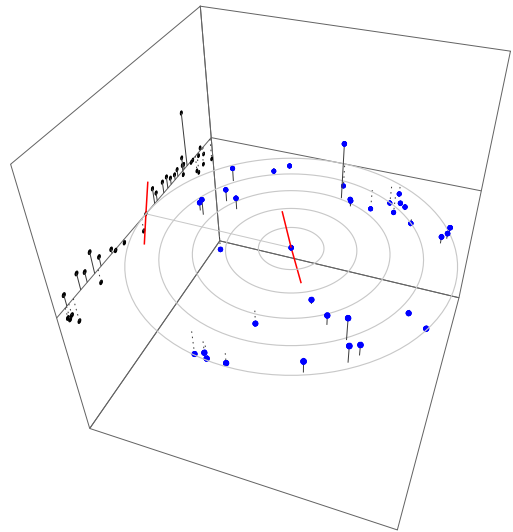}
\caption{
The location and orientation of the polar disk within the wall, between the two voids.  The full volume of the sphere with galaxies brighter than $m_g=17.76$ out to 10 Mpc has been plotted, with concentric circles every 2 Mpc in the plane of the wall. This demonstrates the loneliness of our galaxy and the emptiness of the bounding voids. An edge-on view is projected on the left, showing the thinness of the wall. The red line indicates the position and orientation of the projected major axis of the H {\sc i} disk.
\label{plane}}
\end{figure}

Cosmologically, the galaxy is relatively isolated within a wall between two voids (Fig. \ref{plane}).  A cursory look at 18 other polar ring galaxies taken from \cite{1990AJ....100.1489W} which are contained within the SDSS redshift survey finds them in average or slightly overdense environments (Platen, private communication). This is consistent with a photometric statistical study by \cite{1997A&A...326..907B} of the visible environments of 56 polar ring galaxies, which found them to be similar to normal galaxies.

The two voids and their dividing wall were identified with the closely related Watershed Void Finder (WVF, \citealt{platen2007}) and Cosmic Spine formalism \citep{aragon2008}. The starting point is the Delaunay tessellation field estimator reconstructed density field inferred from the local SDSS galaxy distribution \citep{schaapwey2000}. The WVF identifies the voids and traces their spatial outline, the Cosmic Spine formalism recovers the wall forming the boundary between the two voids. 

Following its identification, we determine the geometry, shape and 
orientation of the wall. These are inferred from the principal 
components of the covariance matrix $\mathcal C_{ij}=\hbox{\rm Var}(r_i r_j)$, computed on the basis of the 34 SDSS galaxies within a distance of $8.5$ Mpc around 
the void galaxy. The wall's flattening is estimated from the 
ratio of its smallest eigenvalue, $e_3$, over the other two eigenvalues 
$e_1$ and $e_2$. The obtained thickness of the wall, $\sim 1.3$ Mpc, 
corresponds to a flattening of $\sim 96\%$.

Fig. \ref{plane} shows the spatial distribution of the galaxies in and around the wall, with the nearest neighbor 4.7 Mpc away, and the boundary voids roughly 25 Mpc in radius.  The size of the upper void is a lower limit as it continues past the edge of the SDSS survey area.  These maps are created using galaxy redshifts from SDSS DR6, which is limited to galaxies with $r<17.77$ magnitudes and 55\arcsec separation from any neighbors.  The polar disk is at an angle of $\sim25^\circ$ from the normal of the wall, and the central stellar disk is similarly misaligned $\sim25^\circ$ away from the plane of the wall.  Roughly, however, the polar disk appears more perpendicular to the wall than aligned.

\section{Discussion} \label{dis}

With our data we can probe in H {\sc i} an area of 650 kpc $\times$ 650 kpc on the plane of the sky surrounding SDSS J102819.24+623502.6, with a velocity range from 4232 km s$^{-1}$ to 6196 km s$^{-1}$.    Despite the significant volume probed, no companions are detected above the $2.0 \times 10^8 M_{\sun}$ limit in H {\sc i}.  Down to the SDSS limits, the neighborhood surrounding our lonely galaxy is empty within 3.5 Mpc.  This environmental isolation makes it unlikely that the substantial $3 \times 10^9 M_{\sun}$ of neutral hydrogen we detect was accreted through a merger or close encounter, and we propose cold accretion as the most likely mechanism \citep{1977ApJ...215..483B, 2005MNRAS.363....2K,2006MNRAS.368....2D}.

Furthermore, satellite mergers are generally accompanied by a burst of star formation and the addition of existing stars to the system, neither of which are seen.  We do see NUV emission in the stellar disk (Fig. \ref{galex}), indicating star formation occurred there within the recent past, however the lack of UV or optical emission in the polar disk is convincing.  Perhaps we are observing this system in the earliest stage of its development, where the slowly accreted gas has not yet cooled sufficiently to begin forming stars.  The measured column density never passes above the $10^{21}$ cm$^{-2}$ surface density threshold proposed by \cite{1987NASCP2466..263S} even without correcting for inclination, leading us to expect little to no star formation in our extended disk.

\begin{figure}[ht!]
\begin{center}
\includegraphics[width=2.7in]{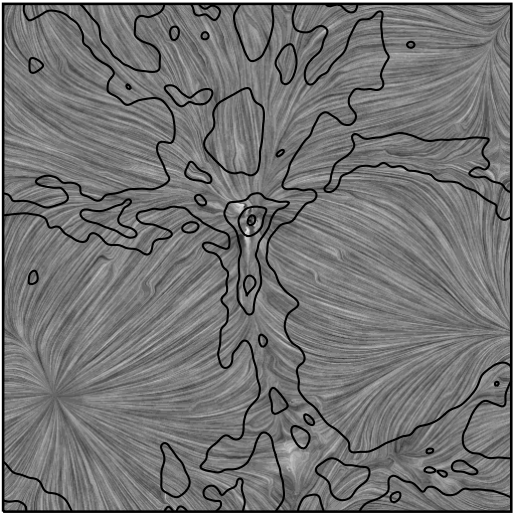}
\caption{N-body simulation showing the flow lines of the cosmic velocity field, with density contours superimposed. Clearly visible is the flow out of the minima onto the wall-like boundaries, and then along these structures towards the high-density peaks. \label{velgrad}}
\end{center}
\end{figure}

Slow accretion has been proposed as a formation mechanism for the optically discovered, H {\sc i} rich polar disk galaxy NGC 4650A \citep{2006ApJ...643..200I}.  While tidal accretion simulations by \cite{2003A&A...401..817B} can reproduce massive, gas-rich and star-poor polar rings, \cite{2005A&A...437...69B} note that such a encounters would convert the central galaxy into an elliptical remnant, and not the rotationally supported disk found in NGC 4650A and suggested in our system. Cold flow accretion as a formation mechanism of polar disks has also been ``observed" serendipitously in two simulations.  \cite{2006ApJ...636L..25M} discovered a polar disk galaxy that formed in their simulation area through the infall of cold gas along a 1 Mpc long filament.  \cite{2008arXiv0802.1051B} found that their cosmological simulation contained a polar disk galaxy that was stable over timescales of 3-5 Gyr.  In both cases, comparisons of the simulated galaxies with observations produce amazing optical and dynamical similarities.

The orientation of the central stellar disk $\sim25^\circ$ from the wall is surprising.  Recent observational and simulation results \citep{2006ApJ...640L.111T,2007MNRAS.375..184B,2007ApJ...655L...5A, 2007MNRAS.381...41H} show that disk galaxies statistically align their spin axis to lie within the wall they are embedded in.  Such an alignment would then naturally lend itself to accretion of material within the plane of the wall, forming a polar disk.  This, however, is not what we see, and instead the accreted material appears to be falling in out of the voids.

What we may witness reflects the process of a hierarchically 
evolving void population \citep{2004MNRAS.350..517S}. As voids 
expand and evacuate matter from their interior, they tend to 
merge with neighboring voids into ever larger underdensities. The 
heuristic simulation of \cite{1993ApJ...410..458D} illustrates 
how a group of spherical voids, embedded within a larger 
underdense area, expand until they meet. 
Once they touch, matter at the sheetlike boundaries of the voids starts to 
stream along these interstices towards the emerging higher density boundary 
surrounding the entire complex of merging voids (see Fig. \ref{velgrad}). As a result we observe the 
thinning and dissolution of the intervoid walls while they get gradually 
integrated in the emerging larger-scale void and remain visible as its tenuous 
interior substructure. The position of our galaxy within its 
wall is reminiscent of this situation, supporting the idea that the 
polar gas is not falling in from the wall, as that material is 
instead streaming away, but out of the bounding voids.

We have discovered completely by chance, as in simulations, a key example of polar disk formation in which accretion of cold gas is the most likely formation scenario.

\acknowledgments
We thank the anonymous referee for constructive comments which improved this paper, Tom Oosterloo for his help with the WSRT data and observations and Ching-Wa Yip for providing the spectral analysis parameters.  This research was partly supported by an NSF grant 0607643 to Columbia University, and the Gordon and Betty Moore foundation.  We are all grateful for support from a Da Vinci professorship at the Kapteyn Institute.

\end{document}